\documentclass[nopubldata]{lpr2}
\usepackage{verbatim}
\usepackage{amsmath}

\newcommand{\eps}{\varepsilon}

\begin{document}
\begin{sloppy}
\titlefigure{soliton_excitation_graphene_stack_2}
\abstract{Nonlinear properties of a multi-layer stack of graphene sheets are studied. It is predicted that such a structure may support dissipative plasmon-solitons generated and supported by an external laser radiation. Novel nonlinear equations describing spatial dynamics of the nonlinear plasmons driven by a plane wave in the Otto configuration are derived and the existence of single and multi-hump dissipative solitons in the graphene structure is predicted. }

\title{Dissipative plasmon-solitons in multilayer graphene}
%
\titlerunning{Dissipative plasmon solitons in multilayer graphene}

\author{Daria A. Smirnova\inst{1,*}, Ilya V. Shadrivov\inst{1}, Alexander I. Smirnov\inst{2}, and Yuri S. Kivshar\inst{1}}

\authorrunning{D. A. Smirnova et al.}

\institute{Nonlinear Physics Center, Research School of Physics and Engineering, 
Australian National University, Canberra ACT 0200, Australia
\and
Institute of Applied Physics, Russian Academy of Sciences, Nizhny Novgorod 603950, Russia
}

\mail{\textsuperscript{*}\,Corresponding author: e-mail: daria.smirnova@anu.edu.au}
%
%
\maketitle

Graphene plasmonics is a rapidly growing new field of physics which utilizes concepts of conventional metal plasmonics combined with unique electronic and optical properties of graphene~\cite{JablanPRB,RevGrigorenko,RevBao,JablanReview,RevLuo}. Similar to metals, graphene can guide electromagnetic waves propagating along its surface due to the coupling of the electromagnetic field to the electron excitations. At the same time, electromagnetic properties of graphene differ significantly from those of metals, therefore graphene plasmon-polaritons may exist in different frequency ranges, and they have substantially different properties. Surface plasmons supported by a graphene monolayer have extremely short wavelength~\cite{JablanReview} and their excitation is quite challenging. Nevertheless, recent experiments ~\cite{Basovexp,Koppens_exp} demonstrated that it is possible to excite and observe such plasmons using scattering near-field scanning optical microscopy.

In the context of potential optical and plasmonic applications, graphene can be incorporated into various photonic structures, including many waveguiding components. It can be placed on top of a variety of substrates, both dielectrics and metals, or it can be even suspended. However, the coupling of electromagnetic energy to graphene in such integrated systems is extremely difficult because of a large wavenumber mismatch. In order to decrease the wavenumber of the waves guided by graphene, one can employ multilayer graphene, a structure consisting of a few graphene monolayers with small enough separations between the layers~\cite{Aleshkin}, whose effective conductivity increases and the wavenumber of the surface plasmons substantially decreases with the number of layers. The corresponding wavenumber reduction is accompanied by a weaker mode localization, which is important for inter-component coupling. 
This suggests that, in such structures, losses of the electromagnetic energy will be relatively low, and the propagation length of plasmons can be increased substantially as compared to a single layer graphene.

In this paper, we consider the generation of surface plasmons in the presence of a plane wave incident from a high-refractive-index dielectric material at the angle larger than the angle of the total internal reflection, see Fig.~\ref{fig1}(a). Surface plasmons are supported by a graphene multilayer being coupled to the incident plane wave via an evanescent field. We demonstrate that in such a geometry the plane wave can act as an external driving source for the plasmon excitations, or as a pump that compensates the propagation losses. The plane wave can resonantly excite surface plasmons and, in the nonlinear case, dissipative plasmon-solitons. We extend the previous results for the soliton formation in a lossless monolayer presented in Ref.~\cite{LPR} to the non-conservative case when the self-localized waves are {\em dissipative solitons}, in which not only nonlinearity has to compensate diffraction, but also gain has to compensate loss. For such solitons no analytical expression can be found, and the soliton profiles have to be calculated numerically.

As suggested in Refs~\cite{Hanson, Aleshkin}, for the multilayer graphene the conductivity of the $N$-layer structure is $N$ times larger than that of the monolayer graphene. The validity of this assumption depends on the interlayer hopping and a type of stacking~\cite{Nakamura_PRB}. However, as was established experimentally, epitaxial non-Bernal stacked multilayer graphene exhibits optical properties similar to those of a single layer~\cite{Orlita_Review}, and these properties are defined by Dirac-like electrons. Furthermore, it can be shown that the dispersion relation for the lowest order electromagnetic mode guided by $N$ closely spaced graphene layers approaches at low frequencies the dispersion of a monolayer graphene, for which one assumes the conductivity $N$ times larger than that of graphene. In particular, for $N=2$ this follows from the dispersion relation of a symmetric TM mode supported by a double-layer graphene waveguide considered in Ref.~\cite{buslaev_JETP}.

In the semiclassical limit, which can be applied in THz and lower frequency ranges, the linear conductivity of graphene can be reduced to the Drude form, which takes into account intraband processes and includes finite relaxation time, usually estimated from the measured impurity-limited DC mobility. The Drude model gives the conductivity in the form~\cite{Mikh_Nonlin, falk}
$$\sigma_{\text{intra}} (\omega)  =  \displaystyle{\frac{ie^2}{\pi \hbar^2}\frac{\mathcal{E}_{F}}{\left(\omega + i\tau_{\text{intra}}^{-1}\right)}},$$
where $\mathcal{E}_{F}$ is the Fermi energy, and $\displaystyle{\tau_{\text{intra}}^{-1}}$ is the relaxation rate. This model is applicable in low temperature limit (i.e., $k_{B}T \ll \mathcal{E}_{F}$) for the relatively low photon energies, i.e. low frequencies, ($\hbar \omega \leq \mathcal{E}_{F}\:$). For high enough field intensities, graphene conductivity becomes nonlinear, and the so-called self-action term or nonlinear correction to the graphene conductivity $\sigma^{\text{NL}}$, was obtained analytically within the quasi-classical approach based on the solution of the kinetic Boltzmann equation for massless quasiparticles in the collisionless approximation~\cite{Mikh_Nonlin, Mikh_Ziegler_Nonlin, Glazov2013}.

The thickness of a stack of graphene layers that we consider here is much smaller than both the plasmon wavelength and transverse confinement of a surface plasmon. This condition is satisfied even for the number of layers up to $20$ with the separations between layers of several nanometers. As a result, a stack of layers can be treated as a conductive sheet described by the Dirac $\delta$-function, similar as in the case of graphene monolayer~\cite{Falk_Persh}. Thus, the electrodynamic formalism for both mono- and multi-layer structures is the same, and in what follows we derive equations applicable to both types of the structures.

We consider a planar geometry shown in Fig.~\ref{fig1}, with a conductive sheet placed at $x=-d/2$. The sheet is surrounded by a dielectric material with permittivity $\eps$. In the numerical examples below, without loss of generality, we assume that $\eps = 1$. A high-dielectric-index material with dielectric constant $\eps_h$ occupies the half-space $x>d/2$. Plane electromagnetic wave is incident on the interface between high and low dielectric index materials. We suppose that the plane wave is incident at the angle greater than the angle of the total internal reflection, and with the tangential component of the wavevector close to the plasmon wavenumber. As a result, the surface plasmon wavenumber $\beta_1$ normalized to $\omega/c$, satisfies the condition $\sqrt{\eps_h}>\beta_1>\sqrt{\eps}$.  The distance between the dielectric halfspace and graphene $d$ is large enough so that its coupling to the plane wave is exponentially weak, and we use this assumption in the asymptotic model developed below.


%
\begin{figure} [t] 
\centerline{\mbox{\resizebox{8.5cm}{!}{\includegraphics{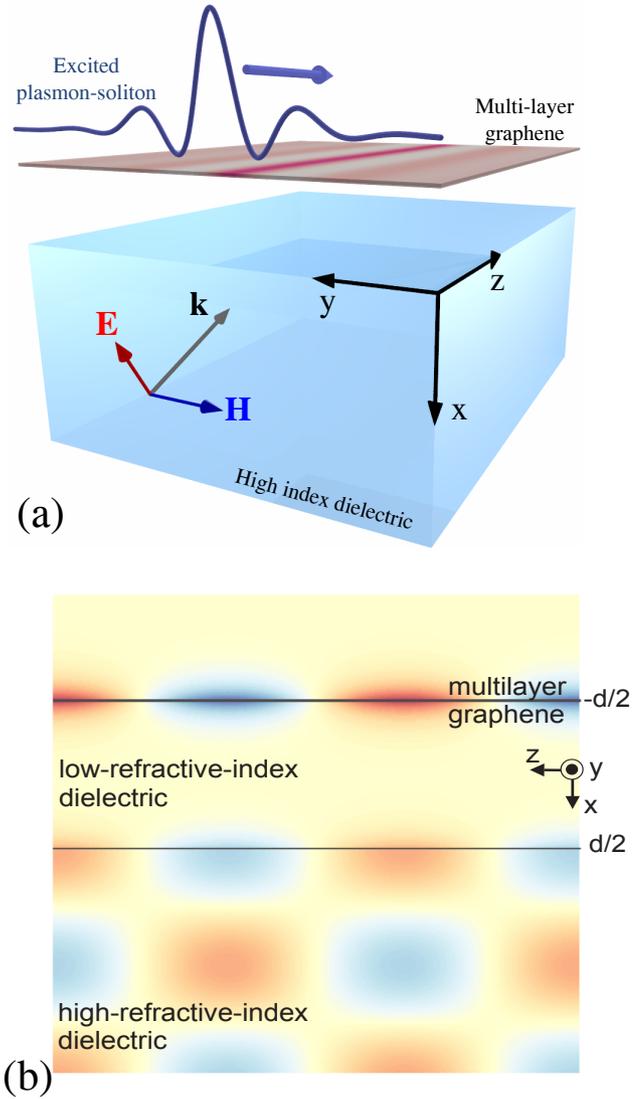}}}}    
\caption{\label{fig:1} (Color online)
(a) Schematic of the problem. (b) Side view of the structure overlapped with profile of the $z$-component of the electric field. Wave incident from the right bottom corner is reflected from the interface, and forms an interference pattern for $x>d/2$. Due to evanescent coupling it excites surface plasmon wave guided by the graphene structure at $x=-d/2$.  }
\label{fig1}
\end{figure}

We start our derivation from the Maxwell's equations written in the form
\begin{equation}
\left \{
\begin{aligned}  \label{eq:eqMaxw1}
\nabla \times {\bf E}&=ik_{0} {\bf H}\:,\\
\nabla \times {\bf H}&=-ik_{0} \varepsilon_2 (x) {\bf E}+\frac{4\pi }{c} \delta (x+d/2)\hat{\sigma }{\bf E}_{\tau } \:.\\
\end{aligned}
\right.
\end{equation}
Here, ${\bf E} \cdot e^{-i\omega t}$ and ${\bf H} \cdot e^{-i\omega t}$ are the electric and magnetic fields, respectively, $k_0 = \omega/c$ is the wavenumber in free space, $\omega$ is the angular frequency, $c$ is the speed of light, $\hat{\sigma }= N \hat{\sigma_1}$ is an equivalent surface conductivity of the $N$-layer graphene, with each layer characterized by a surface conductivity $\hat{\sigma_1}  = \sigma_{\text{intra}}  + \sigma^{\text{NL}} |E_{\tau}|^{2} \equiv   i \sigma^{(I)} +  \sigma^{(R)}  + \sigma^{\text{NL}}|E_{\tau}|^{2}$. ${\bf E}_{\tau }$ is the tangential component of the electric field, nonlinear conductivity \cite{Mikh_Nonlin, Mikh_Ziegler_Nonlin} is
$$\sigma^{\text{NL}} = -i \displaystyle{\frac{3}{8} \frac{e^2}{\pi \hbar^2} \left(\frac{ e V_{F} } { \mathcal{E}_{F} \omega}\right)^2 \frac{ \mathcal{E}_{F} } { \omega}},$$
where $V_{F}\approx c/300$ is the Fermi velocity, subscript $\tau$ refers to the field component tangential to the interface, and distribution of the dielectric permittivity is described by a step function
\begin{equation}
\varepsilon_2(x)=
 \begin{cases}
 \varepsilon\:,&\text{} x<d/2,\\
  \varepsilon_h\:,&\text{} x>d/2.
 \end{cases}\\
\end{equation}
We use the so-called split-field method, and represent the electric and magnetic fields as a superposition ${\bf E} = {\bf E}_{1} + {\bf E}_{2}$, ${\bf H} = {\bf H}_{1} + {\bf H}_{2}$ satisfying the following set of equations:
\begin{align} \label{eq:eqMaxw2}
& \nabla \times {\bf E}_{1,2}=ik_{0} {\bf H}_{1,2}\:,\nonumber\\
& \nabla \times {\bf H}_{1}=-ik_{0} \varepsilon_1(x) {\bf E}_{1}+\frac{4\pi }{c} \delta (x+d/2)\hat\sigma ({\bf E}_{1 \tau } +  {\bf E}_{2 \tau })\:,\nonumber\\
& \nabla \times {\bf H}_{2}=-ik_{0} \varepsilon_2(x) {\bf E}_{2}-ik_{0}(\varepsilon_2(x)-\varepsilon_1(x)){\bf E}_{1}\:,
\end{align}
where $\varepsilon_1(x) = \eps$ is constant. For large spacing $d$ the fields with index "1" describe surface plasmon polaritons supported by graphene layer, while the fields with index "2" describe wave scattering by the interface between $\varepsilon$ and $\varepsilon_h$.

Following the approach employed earlier in Refs.~\cite{Gorbach2013, Coupler_PRB}, we assume that dissipation is low, the nonlinear part of conductivity is much smaller than the linear part, waves are localized in the $y$ direction, with localization width $\Lambda$ such that diffraction remains weak ($k_0 \beta_1 \Lambda \gg 1$). We introduce the small parameter
\begin{equation}
\mu^2 = \text{max} \left\{   \left| \dfrac{\sigma^{(R)}}{\sigma^{(I)}}\right|, \ \left| \dfrac{\sigma^{N\! L} |E_{\tau} |^{2}}{\sigma^{(I)}}\right|, \ (k_0 \beta_1 \Lambda)^{-2}, \ e^{-2\kappa d}\right\}\:,
\end{equation}
where $\kappa = k_0\sqrt{\beta_1^2 - \varepsilon}$, so that the radiative losses of the plasmon are small and proportional to $e^{-2\kappa d}$. Equations~(\ref{eq:eqMaxw2}) can now be rewritten as
\begin{equation}\label{eq:eqMaxw3}
\begin{cases}
\nabla \times {\bf E}_{1,2}=ik_{0} {\bf H}_{1,2}\:,\\
\nabla \times {\bf H}_{1}=-ik_{0} \varepsilon_{1}(x) {\bf E}_{1}+ \frac{4\pi }{c} \delta (x + d/2)i\sigma^{(I)} {\bf E}_{1 \tau } + \mathcal F_{1}\:,\\
\nabla \times {\bf H}_{2}=-ik_{0} \varepsilon_2(x) {\bf E}_{2} + \mathcal F_{2}\:,
\end{cases}
\end{equation}
where the interaction terms
\begin{align}
\mathcal F_{1} & =  \frac{4\pi }{c} \delta (x+d/2) \left\{
	\left[ \sigma^{(R)}+ \sigma^{\text{NL}} |E_{1\tau}|^2 \right]{\bf E}_{1 \tau}+ i\sigma^{(I)}{\bf E}_{2 \tau} \right\} ,\nonumber\\
\mathcal F_{2} & =  -ik_{0}\left[\varepsilon_2(x)-\varepsilon_1(x)\right]{\bf E}_{1},
\end{align}
are small perturbations such that $\mathcal F_{1} \sim \mu^2$ and $\mathcal F_{2} \sim \mu$. As discussed above, by setting $\mathcal F_{1}  = 0$ we obtain the equation for the surface plasmons on a graphene layer, whose profile $h_1(x)$ is then written as
\begin{equation} \label{eq:TrStr1}
  h_{1}(x)= ik_0 \left( \dfrac{\varepsilon} {\kappa} \right) e^{-\kappa |x + d/2|}
	\begin{cases}
		1,&\text{}\ x> - d/2\:,\\
		-1,&\text{} x< - d/2\:,
	\end{cases}
\end{equation}
and the dispersion relation for the plasmon is found from the boundary conditions,
\begin{equation}  \label{eq:DispRel1}
\dfrac{2\varepsilon}{\kappa}=\dfrac{4\pi}{\omega} \sigma^{(I)} \:.
\end{equation}

Now we employ the perturbation theory and represent the magnetic field in the form ${\bf H}_{1} = \{{H_{1\:y }, H_{1\:z } }\}$
\begin{equation} \label{eq:ansatz}
\begin{aligned}
H_{1\:y } & = \biggl[ \mathcal{ A }_{1}(z,y)h_{1}(x)+ H_{1}^{(2)}(z,y,x) \biggr]e^{ik_0\beta_1 z}\:, \\
H_{1\:z } & = { H }_{1}^{(1)}(z,y,x)   e^{ik_0\beta_1 z}\:,
\end{aligned}
\end{equation}
where $H_{1\:y,z}^{(1,2)} \sim \mu ^{1,2} $, $ \dfrac{\partial \mathcal A_{1}}{\partial y}  \sim \mu$. These assumptions lead to $\dfrac{\partial \mathcal A_{1}}{\partial z} \sim \mu^2$, since the complex plasmon amplitude changes along $z$ axis due to dissipation, diffraction and radiative effects, which are considered small. Substituting Eq.~(\ref{eq:ansatz}) into Eq.~(\ref{eq:eqMaxw2}), in the first order of $\mu$ we obtain an equation for the field ${\bf H}_{2} = H_{2}^{(1)} {\bf y}_0$ in the form
\begin{equation}
\begin{aligned}
 & \dfrac{d^2 H_{2}^{(1)}}{dx^2} +  k_0^2(\varepsilon_{2}(x) - \beta_{1}^2)H_{2}^{(1)}  \\
 & \quad \quad \quad \quad \quad + i k_0 (\varepsilon_h - \varepsilon){\delta}(x - d/2) E_{2z}^{(1)}=F_{2}\:,\\
 & F_{2} = k_0^2 \left[\varepsilon_2(x) - \varepsilon_{1}(x)\right]\mathcal A_1  h_1(x) \\
 & + ik_0 (\varepsilon-\varepsilon_h){\delta}(x - d/2)\mathcal A_1   e^{-\kappa d}\:.
\end{aligned}
\end{equation}

The plane wave incident from the semi-space $x>d/2$ is written as $H_0e^{-ik_{2}(x-d/2)+ik_0\beta_2z}$, where we assume  $k_{2} \equiv k_0\sqrt{\varepsilon_h - \beta_2^2} \approx k_{h} \equiv k_0\sqrt{\varepsilon_h - \beta_1^2}$ (i.e. $|\beta_{1}-\beta_{2}|/\beta_{1,2}\sim \mu^2$), and $H_0 \sim \mu$. The wave
experiences the total internal reflection from the interface $x=d/2$.
Taking into account the boundary conditions of continuity of the tangential components of the magnetic and electric fields, ${\bf H} =  {\bf H}_{1} + {\bf H}_{2}$, $E_{z} = E_{1\:z} + E_{2\:z}$ at $x=d/2$, we find the amplitude of the electric field at the interface in the following form
\begin{equation}
E_{2z}^{(1)}(d/2) = \frac{\varepsilon k_{h} - i \varepsilon_h\kappa}{\varepsilon k_{h} + i \varepsilon_h\kappa}\mathcal A_1  e^{-\kappa d} - \dfrac{\kappa}{ik_0 \varepsilon} \frac{2H_0}{1 + i {\varepsilon_h\kappa}/{k_{h}\varepsilon}}\:.
\end{equation}

Finally, we return to the second equation of Eqs.~(\ref{eq:eqMaxw2}) and in the second order of $\mu$ obtain an equation for the correction $H_1^{(2)}$:
\begin{equation} \label{eq:cor1}
\begin{aligned}
 & \dfrac{d^2 H_1^{(2)}}{dx^2} - \kappa^2 H_1^{(2)} - \dfrac{4 \pi}{c} i \sigma^{(I)}\dot{\delta}(x + d/2) E_{1z}^{(2)} = F_{1}\:,\\
 & F_{1} = -\left( 2ik_0\beta_{1} \dfrac{\partial \mathcal A_{1}}{\partial z} + \dfrac{\partial^2 \mathcal A_{1}}{\partial y^2}\right)h_1(x)
 + \frac{4 \pi}{c} \dot{\delta}(x + d/2)\cdot \\
 &\left[ \sigma^{(R)} + \sigma^{\text{NL}}|\mathcal A_1  |^2\right] \mathcal A_{1} + i\sigma^{(I)} E_{2z}^{(1)}(d/2) e^{-\kappa d}\:,
\end{aligned}
\end{equation}
where $\dot{\delta}$ is the derivative of the Delta-function.
According to the Fredholm theorem of alternative\cite{Korn}, the solution of Eq.~(\ref{eq:cor1}) is non-divergent provided that the eigenmodes of its homogeneous part are orthogonal to the right-hand-side of the equation, and this condition can be mathematically written as $\displaystyle{\int\limits_{-\infty}^{+\infty}F_1(x)h_1^{*}(x)dx = 0}$.
Applying this theorem to Eq.~(\ref{eq:cor1}), we obtain the equation for the envelope of the surface plasmon polariton in the form
\begin{multline} \label{eq:LL}
2ik_0\beta_1\left(\dfrac{\partial \mathcal A_{1}}{\partial z} + (\gamma + \gamma_{r} ) \mathcal A_{1} + i \Delta_r \mathcal A_{1} \right)  +\dfrac{\partial^2 \mathcal A_{1}}{\partial y^2} {} \\
 {} + g |\mathcal A_1  |^2 \mathcal A_{1} = Q H_0 e^{i\Delta z}\:,
\end{multline}
where linear detuning $\Delta$, linear damping due to losses in graphene $\gamma$, damping due to radiation into the high index medium $\gamma_r$, radiative detuning $\Delta_r$, nonlinear detuning coefficient $g$, coupling coefficient $Q$ are derived as
\begin{gather}
g = \dfrac{4\pi}{c} \sigma^{\text{NL}} \dfrac{ i k_0 \varepsilon}{q}\:, \quad \gamma = \dfrac{2 \pi}{c \beta_1} \sigma^{(R)}\dfrac{\varepsilon}{q}\:, \nonumber   \\
Q = \dfrac{4\pi}{c} i \sigma^{(I)} \dfrac{ \kappa}{q}  \frac{2 e^{-\kappa d}}{1 + i \frac{\varepsilon_h\kappa}{k_{h}\varepsilon}} \:,\quad \Delta  = k_0(\beta_2 - \beta_1)  \:, \nonumber \\
R \equiv \gamma_r + i\Delta_r = \dfrac{2 \pi}{c \beta_1} i \sigma^{(I)} \dfrac{\varepsilon}{q}  \frac{\varepsilon k_{h} - i \varepsilon_h\kappa}{\varepsilon k_{h} + i \varepsilon_h\kappa} e^{-2\kappa d} \:, \nonumber\\
q  \equiv \displaystyle{\int\limits_{-\infty}^{+\infty}h_1(x)h_1^{*}(x)dx} = {k_0^2} \dfrac{\varepsilon^2}{\kappa^3}\:.
\end{gather}
The right-hand side of Eq.~(\ref{eq:LL}) contains an external force, which in our further simulations is homogeneous due to plane wave excitation.
\begin{figure} [tb]
\centerline{\mbox{\resizebox{8.5cm}{!}{\includegraphics{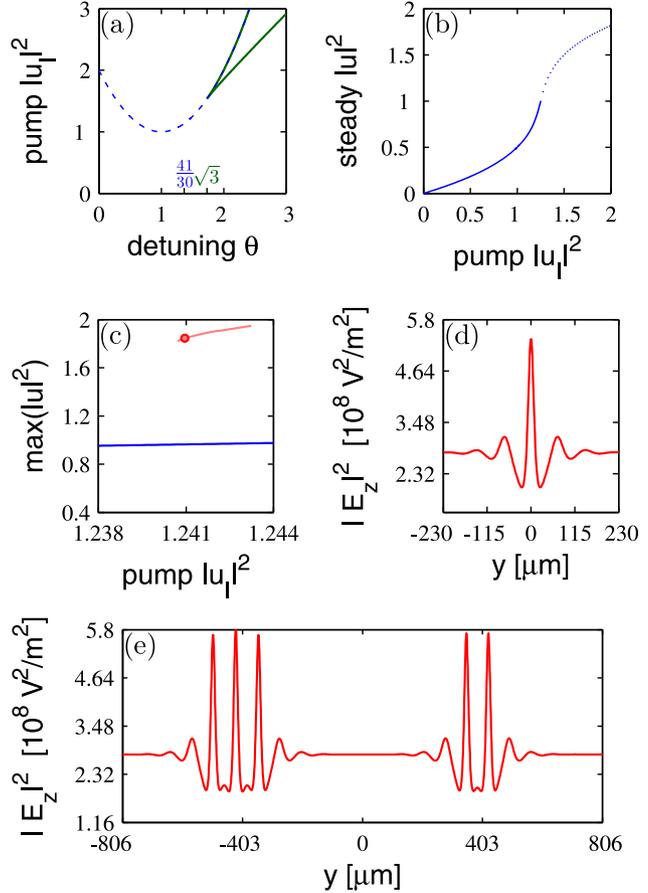}}}}
\caption{\label{fig:2} (Color online)
(a) The plane of parameters of pump intensity and detuning. Two green curves show the boundary of the bistable region: inside the curves there are three equilibrium states, on the lines there are two equilibria, and one equilibrium exists outside this region. The homogeneous excitation becomes modulationally unstable above the dashed line, which is defined by $|u|^2=1$. (b) Intensity of the excited plasmon as a function of pump intensity for $\Theta = 1.5$. The dotted curve indicates modulationally unstable part of the dependence. (c) Single peak soliton properties: upper and lower curves show peak soliton intensity and steady background, respectively, $\Theta = 1.5$.  (d) Transverse profile of a soliton calculated for the parameters corresponding to the dot in the Fig.~\ref{fig3} (c). 
(e) Transverse profile of plasmon-solitons containing two and three peaks.}
\label{fig2}
\end{figure}
\begin{figure} [tb]
\centerline{\mbox{\resizebox{8.5cm}{!}{\includegraphics{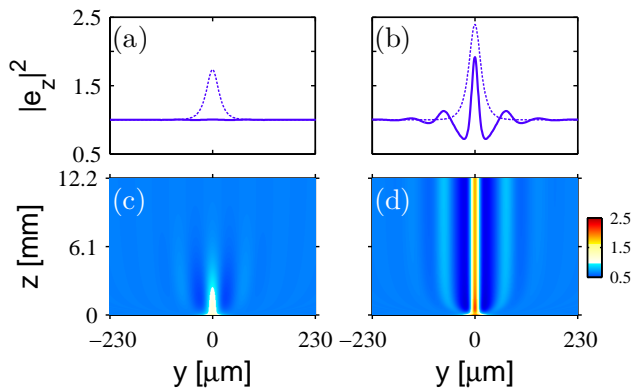}}}}
\caption{\label{fig:3} (Color online)
Evolution of the spatial profile of the localized plasmon excited at $z=0$.
Initial profiles [dashed lines in (a,b)] are of the same widths but of different amplitudes. (a,c) Diffraction and damping of the initial excitation and (b,d) soliton formation. Shown is the intensity normalized to the homogeneous steady state background. In panels (a) and (b) dashed and solid lines correspond to the initial and final intensity distributions, respectively.
}
\label{fig3}
\end{figure}
To simplify our further analysis, we introduce new variables
\begin{gather}
\xi = (\gamma_r + \gamma)z\:, \quad \eta = \sqrt{2 k_0\beta_1 (\gamma_r + \gamma) }y \:, \quad \Theta = \frac{  \Delta_r + \Delta }{\gamma_r + \gamma}  \:,  \nonumber\\
u = \frac {P_1} {|\mathcal A_1  |^2} \sqrt{\frac{2 k_0\beta_1 (\gamma_r + \gamma) }{g}} \mathcal A_1  e^{-i\Delta z}\:, \nonumber \\
P_1= \displaystyle{\int\limits_{-\infty}^{+\infty}\frac{c}{8 \pi} \left[{\bf E}_{1}  {\bf H}_{1}^{*}\right]_z dx = \frac{c}{8 \pi k_0} \frac{\beta_1 \epsilon}{(\beta_1^2 - \epsilon)^{3/2}} |\mathcal A_1  |^2  + O(\mu ^2)   }\:,  \nonumber\\
i u_I =  Q \sqrt{\frac{g}{(2 k_0\beta_1 (\gamma_r + \gamma))^3} } H_0
\end{gather}
and obtain the dimensionless form of Eq.~(\ref{eq:LL})
\begin{equation} \label{eq:LL_norm}
\dfrac{\partial u} {\partial \xi}  =   u_{I} - (1 + i \Theta - i |u |^2 ) u + i\dfrac{\partial^2 u}{\partial \eta ^2} \:.
\end{equation}
This is the Lugiato-Lefever equation~\cite{LL}, studied in a number of previous works. In particular, bifurcation analysis is available in Ref.\cite{Ackermann} and references therein. Here, we only briefly focus on some of the features of this equation, which are important for the effects related to our plasmon-soliton excitation. For the steady state homogeneous solutions of Eq.~(\ref{eq:LL_norm}), we obtain a cubic equation. This equation has only one solution when $\Theta<\sqrt{3}$, and three solutions otherwise, indicating the presence of bistability, see Fig.~\ref{fig2}(a). The latter domain of parameters comprises several bifurcations that, in particular, may result in {\em breathing solitons} and even {\em chaos}~\cite{Turaev, breathers_oe_2013}. If $\Theta<\sqrt{3}$, the only bifurcation that occurs leads to the Turing patterns~\cite{LL}. At the critical value of detuning $\Theta = 41/30$, the character of this bifurcation changes from supercritical to subcritical. In subcritical regime, the stationary solutions have the form of localized dissipative structures (usually called {\em cavity solitons}) and their bound states sitting on the background of a homogeneous steady state solution~\cite{Gomila}. Considering the case when the latter is a single-valued function of the pump, in Figs~\ref{fig2}(c,d) we show the dependence of the soliton amplitude on the pump intensity and also the soliton profile. This soliton has a single main intensity peak falling away into oscillating tails. Soliton is calculated for the following parameters $\varepsilon = 1$, $\varepsilon_h = 11.56$, the Fermi energy $\mathcal{E}_{F}=0.2$ eV, $\hbar \omega = 0.25 \mathcal{E}_{F}$ ($\lambda=2\pi/k_0\approx 25$ $\mu$m), $\beta_1 = 3.03$,
the relaxation time $\tau_{\text{intra}} = 1$ ps, the number of graphene layers $N = 6$, linear detuning $\beta_2 - \beta_1= 0.09$, 
and $d=3$ $\mu$m. For this structure and considered wave intensities, the smallness parameter $\mu \sim 0.1$ and our approximate approach is well justified. To put the spatial scales in some context, the ratios of the plasmon wavelength to soliton width, to free space wavelength and to the Fermi wavelength are 0.08, 0.33 and 400, respectively. 

We would like to note that if the relaxation time is decreased to 0.1ps, which is what observed in most of the available graphene samples, then in order to balance the increased dissipative losses it is necessary to increase the electric field amplitude in the incident wave as well as radiative losses, roughly, by a factor of 10. This will lead to the decrease of the soliton width by $\sqrt{10}$. However, though the formation of dissipative structures is expected even in this case, the developed asymptotic description is not strictly applicable.

The Lugiato-Lefever equation~(\ref{eq:LL_norm}) is characterised by a snaking bifurcation diagram for cavity solitons~\cite{Ackermann, Gomila}, and that is why the patterned multi-peak solutions can coexist on a stable uniform pedestal. These solutions are frequently regarded as controllable bits for all-optical memory applications\cite{Ackermann}. Noticeably, the plethora of possible steady states can be interpreted as the parts of subcritical Turing patterns with a number of elements defined by initial conditions. Such multi-peak structures with finite even or odd numbers of peaks [see two and three-peak steady solutions in Fig.~\ref{fig2}(e)] are sometimes called soliton molecules or soliton clusters, emphasizing their spatial localization. Importantly, because of the subcritical nature, a bright soliton can not be switched on by an arbitrarily small perturbations of the background field (intensity) and hence low-energetic excitations will decay, as illustrated in Figs~\ref{fig3}(a,c), while Figs~\ref{fig3}(b,d) display the dynamic formation of a soliton from a narrow perturbation of a higher amplitude.

In conclusion, we have derived novel nonlinear equations describing the excitation of dissipative TM-polarized plasmon solitons in graphene. We have employed the multilayer graphene structure in order to increase the wavelength of the plasmon-polaritons to be excited in the Otto configuration for the attenuated total internal reflection. We have obtained the Lugiato-Lefever type nonlinear equation and predicted the existence of single- and multi-peak dissipative solitons in such graphene structures.

\begin{acknowledgement}
The work was supported by the Australian Research Council. A.I.S. acknowledges support from RFBR through grant 13-02-00881.
\end{acknowledgement}

\end{sloppy}
\end{document}